\documentclass[preprint,showpacs,aps]{revtex4}

\usepackage{graphics}

\begin{document}

\title{Vacuum Structure of Two-Dimensional $\phi^4$ Theory \\ on the Orbifold $S^{1}/Z_{2}$}

\author{H. T. Cho}
  \email{htcho@mail.tku.edu.tw}
\affiliation{Department of Physics, Tamkang University, Tamsui,
Taipei, Taiwan, R.O.C.}

\affiliation{Institute of Physics, Academia Sinica, Taipei,
Taiwan, R.O.C.}

\date{\today}

\begin{abstract}
We consider the vacuum structure of two-dimensional $\phi^4$
theory on $S^{1}/Z_{2}$ both in the bosonic and the supersymmetric
cases. When the size of the orbifold is varied, a phase transition
occurs at $L_{c}=2\pi/m$, where $m$ is the mass of $\phi$. For
$L<L_{c}$, there is a unique vacuum, while for $L>L_{c}$, there
are two degenerate vacua. We also obtain the 1-loop quantum
corrections around these vacuum solutions, exactly in the case of
$L<L_{c}$ and perturbatively for $L$ greater than but close to
$L_{c}$. Including the fermions we find that the ``chiral" zero
modes around the fixed points are different for $L<L_{c}$ and
$L>L_{c}$. As for the quantum corrections, the fermionic
contributions cancel the singular part of the bosonic
contributions at $L=0$. Then the total quantum correction has a
minimum at the critical length $L_{c}$.
\end{abstract}


\maketitle

\section{Introduction}

The vacuum structure and the soliton solutions of a field theory
can be changed dramatically in the presence of compact dimensions.
A well-known example is that of the Hosotani mechanism
\cite{hosotani} in which vanishing field strength does not
necessarily imply vanishing gauge potential in non-simply
connected spaces. Then the non-vanishing gauge potential can
signify the breaking of gauge symmetries. This is actually related
to the fact that on non-simply connected spaces fields can have
different, or twisted, boundary conditions compatible with the
gauge symmetry \cite{scherk}. On the other hand, one can also
study the dependence of the vacuum structure on these boundary
conditions in the situations with global symmetries only
\cite{hatanaka}.

The allowed soliton solutions of the theory can depend on these
boundary conditions too. For example, the kink solutions of the
two-dimensional $\phi^4$ theory with a symmetry breaking potential
will disappear when the space is compactified to a circle
\cite{manton}, that is, when periodic boundary condition is
imposed on the scalar field. They are replaced by sphaleron
solutions consisting of kink and anti-kink pairs. Since the space
is compact, finite energy requirement no longer presents a
constraint on the possible soliton solutions. Consequently, the
topological classifications of these solutions have to be modified
accordingly.

In addition to compactifying to a circle, one can consider that of
an orbifold like $S^{1}/Z_{2}$. This is related to the equivalence
of translations as well as reflections or parity operations in the
internal dimensions \cite{quiros}. Recently, there are quite a lot
of interests on the construction of GUT models with orbifold extra
dimensions. This is a simple way to obtain chiral zero mode
fermions on the fixed points where the physical dimensions reside
\cite{georgi}. The Hosotani mechanism can also be realized on
compact spaces like the orbifolds \cite{hosotani2}, making it
possible to have symmetry breaking without the Higgs field in
these models.

In order to have a more detailed understanding of the properties
of the scalar as well as the fermion fields on the orbifold, we
consider the simple case of the two-dimensional $\phi^4$ theory in
this paper. In this model most of the analysis can be carried out
explicitly. On the other hand, the results that we obtain here
should also be relevant to the theories in higher dimensions. In
the next section, we consider the vacuum solutions and their
quantum corrections on the orbifold $S^{1}/Z_{2}$ with only scalar
fields. In Section III, we include fermions by introducing a
supersymmetric Lagrangian. The fermionic contributions to the
quantum corrections are then calculated. Conclusions and
discussions are given in Section IV.

\section{Two-Dimensional $\phi^4$ theory}

In this section we consider the $\phi^4$ theory in
(1+1)-dimensions,
\begin{equation}
{\cal
L}=\frac{1}{2}(\partial_{\mu}\phi)^2-U(\phi),\label{lagrangian}
\end{equation}
where
\begin{equation}
U(\phi)=\frac{\lambda}{4}\left(\phi^{2}-\frac{m^{2}}{\lambda}\right)^2.
\end{equation}
First we derive the vacuum solutions on $S^{1}/Z_{2}$ from the
static solutions on $S^1$ \cite{manton} for different scales of
the spatial dimension. Then the quantum corrections about these
solutions are calculated using direct mode sums with zeta-function
regularization \cite{elizalde}.

\subsection{Vacuum solutions}

\begin{figure}[!]
\includegraphics{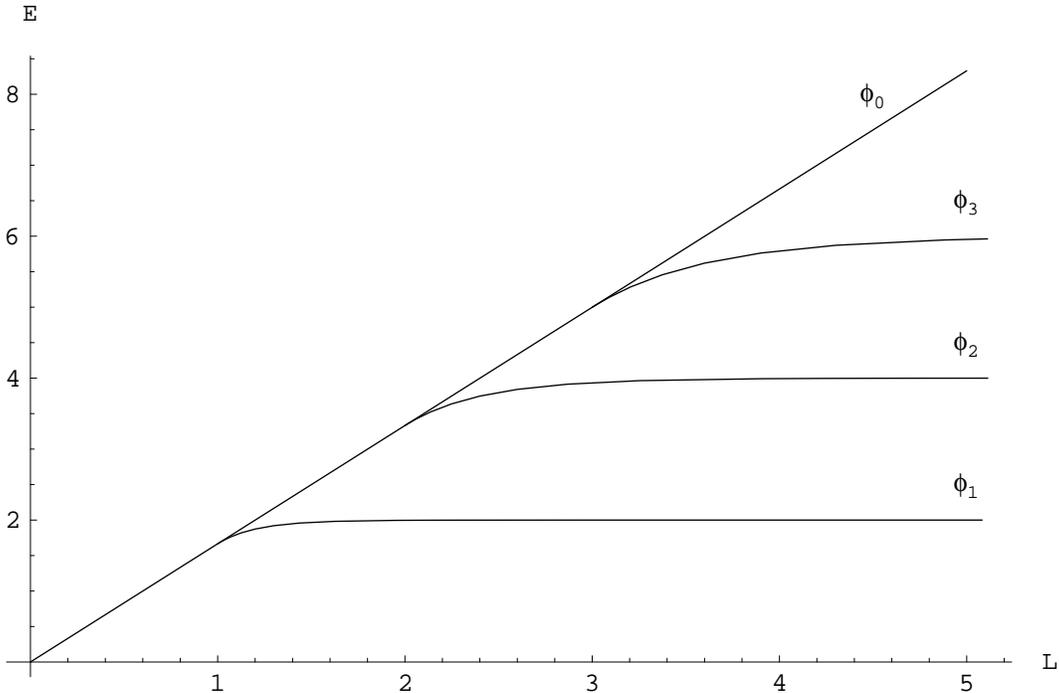}
\caption{\label{orbifoldfig1} Variation of the energies (in units
of the kink energy $E_{0}=2\sqrt{2}m^{3}/3\lambda$) of the static
solutions with the size (in units of the length $L_{1}$) of the
circle.}
\end{figure}

The equation of motion to the Lagrangian in Eq.~(\ref{lagrangian})
is
\begin{equation}
-\partial^{2}\phi-U'(\phi)=0,
\end{equation}
and if we concentrate on the static solutions, we have
\begin{equation}
\frac{d^{2}\phi}{dx^{2}}=-(-U'(\phi)).
\end{equation}
This is just the Newton's second law with $x$ identified as
``time". On the circle $S^{1}$, the solutions \cite{manton} can be
readily obtained if one imposes the periodic boundary conditions,
$\phi(x+L)=\phi(x)$, where $L$ is the perimeter of the circle.
They are the vacuum solutions,
\begin{equation}
\phi_{v}=\pm \frac{m}{\sqrt{\lambda}},
\end{equation}
the unstable solution,
\begin{equation}
\phi_{0}=0,
\end{equation}
and the periodic solutions,
\begin{equation}
\phi_{n}=\frac{m}{\sqrt{\lambda}}\left(\sqrt{\frac{2k^{2}}{1+k^{2}}}\right){\rm
sn}\left[\frac{m}{\sqrt{1+k^{2}}}x\right],
\end{equation}
where sn is the Jacobi elliptic function. $\phi_{n}$ consists of
$n$ pairs of kink and anti-kink. Here, the relation between $L$,
$n$, and the modular parameter $k$ ($0\leq k\leq 1$) is
\begin{equation}
L=\frac{4n\sqrt{1+k^2}}{m}K(k).
\end{equation}
Since the minimum value of $K(k)$ is $K(0)=\pi/2$, the number of
allowed $\phi_{n}$ solutions increases with $L$. For example, the
one kink-anti-kink pair solution exists only when $L\geq
L_{1}=2\pi/m$. On the circle, both $\phi_{0}$ and $\phi_{n}$ are
unstable and they will decay to the vacuum solutions $\phi_{v}$.
The energies of these configurations, in unit of the kink energy
$E_{0}=2\sqrt{2}m^{3}/3\lambda$, as functions of $L$ are plotted
in Fig.~\ref{orbifoldfig1} \cite{manton}. As shown in this figure,
for $L<L_{1}$, $\phi_{0}$ is the only unstable solution of the
theory and it is interpreted as the sphaleron solution. While for
$L>L_{1}$, $\phi_{1}$ is also an unstable solution. Since it has
lower energy than $\phi_{0}$, it becomes the sphaleron solution of
the theory.

On the orbifold $S^{1}/Z_{2}$, one identifies $x$ and $-x$ on the
basic interval $-L/2<x<L/2$. Since the Lagrangian
(Eq.~(\ref{lagrangian})) is invariant under the transformation
$\phi\rightarrow -\phi$, one could require
\begin{equation}
\phi(-x)=\pm\phi(x),
\end{equation}
that is, one could impose either parity on the field. Not much is
changed if one imposes the positive parity condition on the field.
However, if one chooses instead the negative parity condition, the
vacuum structure of the theory can be changed dramatically.
Requiring the field to have the orbifold constraint,
\begin{equation}
\phi(-x)=-\phi(x),
\end{equation}
the vacuum solutions $\phi_{v}$ are excluded because they have
even parities. Then for $L<L_{1}$, $\phi_{0}$ becomes the unique
vacuum solution. For $L>L_{1}$,
\begin{equation}
\phi_{\pm 1
}=\pm\frac{m}{\sqrt{\lambda}}\left(\sqrt{\frac{2k^{2}}{1+k^{2}}}\right){\rm
sn}\left[\frac{m}{\sqrt{1+k^{2}}}x\right],
\end{equation}
with $k$ implicitly given by $\sqrt{1+k^2}K(k)=mL/4$, are the
degenerate vacuum solutions. Note that because of the orbifold
constraint, which breaks the translational symmetry of the static
solutions, $\phi_{1}$ and $\phi_{-1}$ can no longer be considered
as the same solution. In effect, a phase transition occurs when
$L$ is varied. The critical perimeter is
\begin{equation}
L_{c}=L_{1}=\frac{2\pi}{m}.\label{critical}
\end{equation}

It is interesting to note that this phase transition is
specifically related to the orbifold structure of the space. It
will not occur in the case of a circle with periodic boundary
condition alone. Here for $L<L_{c}$, we have the unique vacuum
$\phi_{0}$. At $L_{c}$, $\phi_{0}$ bifurcates into three
solutions, $\phi_{0}$, $\phi_{1}$, and $\phi_{-1}$. $\phi_{\pm 1}$
become the degenerate vacua and $\phi_{0}$ the sphaleron solution.
When the orbifold is taken as the extra dimension in a higher
dimensional theory, the above phase transition will certainly be
relevant to its vacuum structure. Moreover, if more fields are
added to the theory, their properties will be affected by it too.
We shall discuss the case of fermions in Section III. Before doing
that we shall first consider the stability of these vacuum
solutions and the quantum corrections to them.

The stability of the vacuum solutions can be analyzed by looking
at the perturbations around them \cite{manton}. For $\phi_{0}=0$
with $L<L_{1}$, the perturbation equation is just
\begin{equation}
\frac{d^{2}\eta}{dx^{2}}+(\omega^{2}+m^{2})\eta=0,
\end{equation}
where $\phi(x,t)=\phi_{0}+\eta(x)e^{-i\omega t}$. This is just a
harmonic oscillator equation. Imposing the periodic boundary
conditions, the solutions are simply
\begin{equation}
\eta_{0}\sim {\rm const}
\end{equation}
with the frequency $\omega_{0}^{2}=-m^{2}$, and
\begin{equation}
\eta_{p}\sim{\rm sin}\frac{2\pi px}{L}\ \ ,\ \ {\rm cos}\frac{2\pi
px}{L}\label{eigenfunction}
\end{equation}
with
\begin{equation}
\omega_{p}^{2}=\frac{4\pi^{2}p^{2}}{L^{2}}-m^{2}
=m^{2}\left[\left(\frac{L_{1}}{L}\right)^{2}p^{2}-1\right],\ \
p=1,2,\dots\label{phi0spect}
\end{equation}
The lowest energy state is the only negative mode arising from the
fact that $\phi_{0}$ is unstable and it can decay to the vacuum
solutions. Now, if the orbifold constraint is imposed, this
negative mode will be excluded because it is even. $\phi_{0}$ is
therefore stable and becomes the unique ground state of the theory
for $L<L_{1}$.

Similarly, one can analyze the stability of $\phi_{\pm 1}$ for
$L>L_{1}$. Here the perturbation equation becomes \cite{liang}
\begin{equation}
\frac{d^{2}\eta}{dx^{2}}+(\omega^{2}+m^{2}-3\lambda\phi_{1}^2)\eta=0.
\end{equation}
This is a Lam\'e equation. Its lowest five eigenfunctions in this
case are given by the Lam\'e polynomials, and the rest by the
Lam\'e transcendental functions \cite{arscott}. The lowest energy
state is a negative mode,
\begin{equation}
\eta_{0}(z)={\rm sn}^{2}(z)-\frac{1}{3k^{2}}\left(1+k^{2}+
\sqrt{1-k^{2}(1-k^{2})}\right),
\end{equation}
where $z=mx/\sqrt{1+k^{2}}$, with frequency
\begin{equation}
\omega^{2}=m^{2}\left(1-\frac{2\sqrt{1-k^{2}(1-k^{2})}}
{1+k^{2}}\right)\leq 0.
\end{equation}
This again indicates that $\phi_{\pm 1}$ are unstable on a circle
and they will decay to the vacuum solutions. Note that this
negative mode has even parity. The orbifold constraint will also
exclude this mode and render $\phi_{\pm 1}$ stable. The next state
is the zero mode with
\begin{equation}
\eta_{1}(z)={\rm cn}(z)\ {\rm dn}(z).
\end{equation}
This state is also even. Its presence is related to the
translational (or rotational) symmetry of $\phi_{n}$. The orbifold
constraint will exclude this mode too because the orbifold
constraint also breaks the translational symmetry of $\phi_{n}$.

The other three Lam\'e polynomial states are
\begin{eqnarray}
\eta_{3}(z)&=&{\rm sn}(z){\rm dn}(z)\ \ ;\ \
\omega_{3}^{2}=\frac{3m^{2}k^{2}}{1+k^{2}}\\ \eta_{4}(z)&=&{\rm
sn}(z){\rm cn}(z)\ \ ;\ \ \omega_{4}^{2}=\frac{3m^{2}}{1+k^{2}}
\end{eqnarray}
which are both odd, and
\begin{equation}
\eta_{5}(z)={\rm
sn}^{2}(z)-\frac{1}{3k^{2}}\left(1+k^{2}-\sqrt{1-k^{2}(1-k^{2})}\right)
\end{equation}
which is even, with
\begin{equation}
\omega_{5}^{2}=m^{2}\left(1+\frac{2\sqrt{1-k^{2}(1-k^{2})}}{1+k^{2}}\right).
\end{equation}
The orbifold constraint will also exclude $\eta_{5}(z)$. Although
the rest of the spectrum is not known explicitly, one can see that
the parities of the eigenfunctions have the pattern: even, even,
odd, odd, even, even, ... Hence, exactly half of the spectrum will
satisfy the orbifold constraint.

\subsection{Quantum corrections}

\begin{figure}[!]
\includegraphics{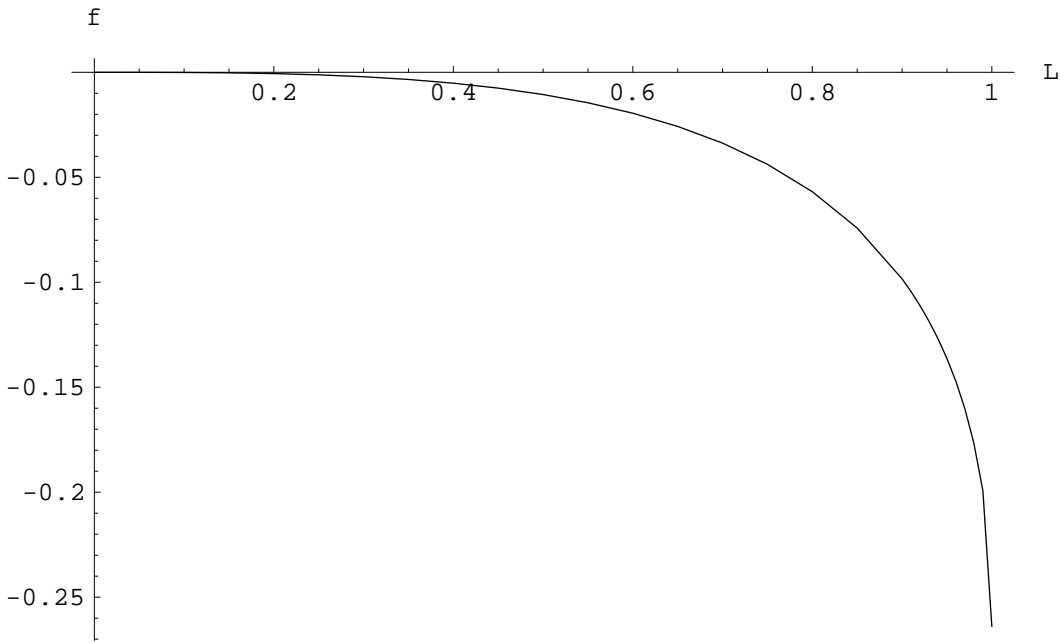}
\caption{\label{orbifoldfig2} The value of the convergent series
$f$ as a function of the size (in units of the length $L_{1}$) of
the compact dimension.}
\end{figure}

\begin{figure}[!]
\includegraphics{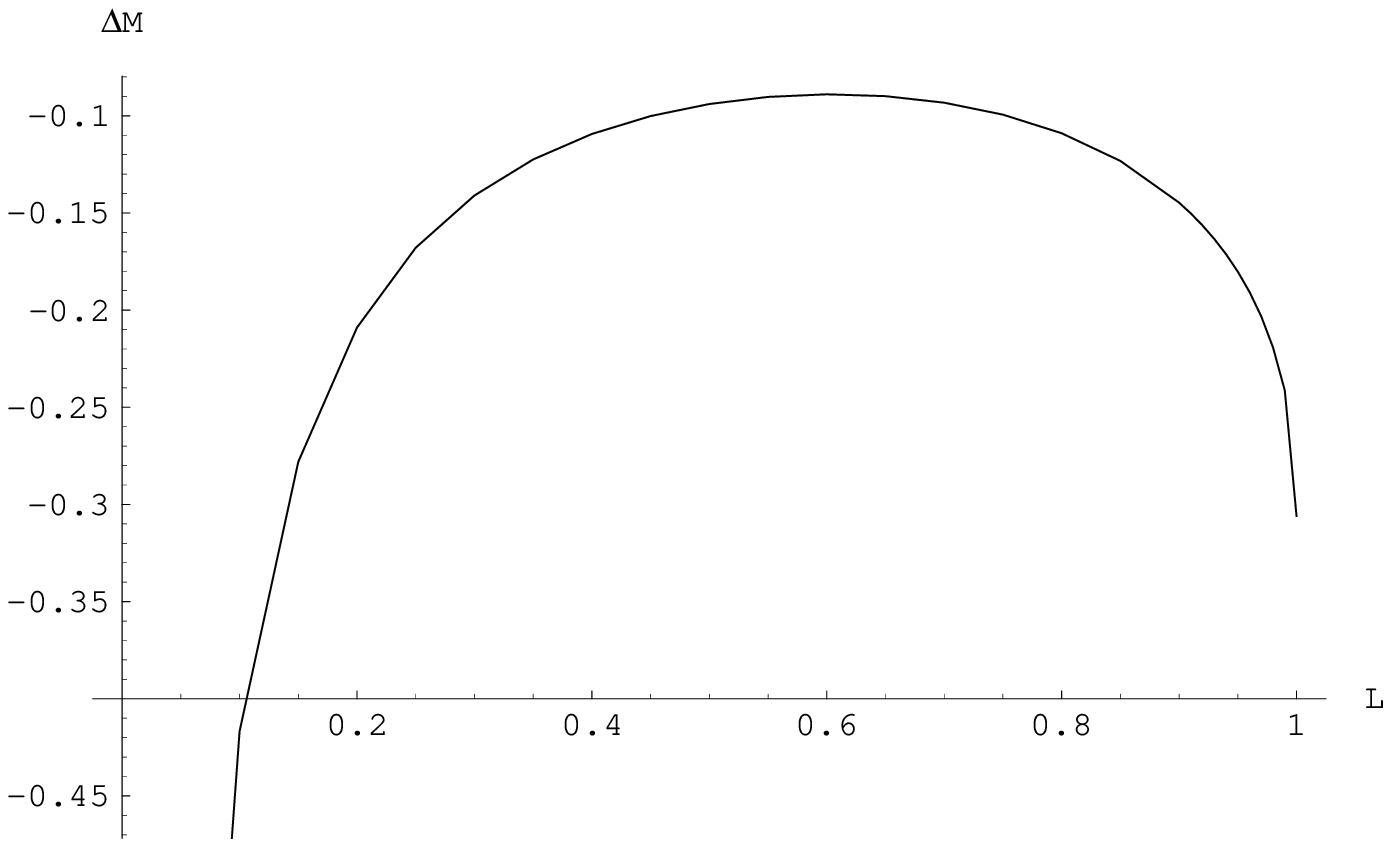}
\caption{\label{orbifoldfig3} The value of the 1-loop quantum
corrections (in unit of the mass parameter $m$) as a function of
the size (in units of the length $L_{1}$) of the compact
dimension.}
\end{figure}

Next we calculate the one-loop quantum corrections to the vacuum
energies. For $L<L_{1}$, the vacuum solution is $\phi_{0}$ with
the classical energy
\begin{equation}
M_{cl}[\phi_{0}]=\frac{m^{4}L}{4\lambda}.
\end{equation}
The quantum correction to this energy can be evaluated explicitly
using the spectrum of perturbations in Eq.~(\ref{phi0spect}),
\begin{eqnarray}
(\Delta M)_{\phi_{0}}&=&\frac{mL_{1}}{2L}\sum_{p=1}^{\infty}
\left[p^{2}-\left(\frac{L}{L_{1}}\right)^{2}\right]^{1/2}\nonumber\\
&=&\lim_{s\rightarrow-1}\frac{1}{2}\left(\frac{mL_{1}}{L}\right)^{-s}
\sum_{p=1}^{\infty}
\left[p^{2}-\left(\frac{L}{L_{1}}\right)^{2}\right]^{-s/2}.
\end{eqnarray}
Here we have used the zeta-function regularization method
\cite{elizalde}. In terms of the Riemann zeta-function, the above
sum can be expressed as
\begin{eqnarray}
(\Delta
M)_{\phi_{0}}&=&\lim_{s\rightarrow-1}\left[\frac{1}{2}\left(\frac{mL_{1}}{L}\right)^{-s}
\zeta(s)+\left(\frac{L}{2L_{1}}\right)^{2}s\left(\frac{mL_{1}}{L}\right)^{-s}
\zeta(2+s)\right.\nonumber\\ &&\ \ \ \ \ \ \ \ \ \left.
+\left(\frac{mL_{1}}{L}\right)^{-s}\sum_{n=2}^{\infty}\frac{(-1)^{n}\Gamma(1-s/2)}
{2\Gamma(n+1)\Gamma(1-n-s/2)}\left(\frac{L}{L_{1}}\right)^{2n}\zeta(2n+s)\right].
\end{eqnarray}
The first term,
\begin{equation}
\frac{1}{2}\left(\frac{mL_{1}}{L}\right)\zeta(-1)=-\frac{\pi}{12L},
\end{equation}
which is just the Casimir energy of a massless scalar field on the
orbifold. This term diverges as $L\rightarrow 0$. The second term
is
\begin{equation}
\lim_{s\rightarrow-1}\left(\frac{L}{2L_{1}}\right)^{2}s\left(\frac{mL_{1}}{L}\right)^{-s}
\zeta(2+s)=\lim_{s\rightarrow-1}\frac{m^{2}L}{8\pi}\left[-\frac{1}{s+1}+1-\gamma+{\rm
ln}\left(\frac{2\pi}{L}\right)\right].
\end{equation}
This term with the pole divergent part can be cancelled by
appropriately choosing the mass renormalization scheme. In fact,
mass renormalization is the only one necessary for the
two-dimensional $\phi^{4}$ theory \cite{dashen}. The last term
gives
\begin{equation}
m\sum_{n=2}^{\infty}\frac{(-1)^{n}\sqrt{\pi}}{4\Gamma(n+1)\Gamma(\frac{3}{2}-n)}
\left(\frac{L}{L_{1}}\right)^{2n-1}\zeta(2n-1)\equiv mf(L/L_{1}),
\end{equation}
where, for $0\leq L\leq L_{1}$, $f(L/L_{1})$ is a convergent
series, which is plotted in Fig.~\ref{orbifoldfig2}, with
$f(1)=-0.264$. Putting all these together, after mass
renormalization,
\begin{equation}
(\Delta M)^{ren}_{\phi_{0}}=-\frac{\pi}{12L}+mf(L/L_{1}),
\end{equation}
which is plotted in Fig.~\ref{orbifoldfig3}. Here we have
\begin{equation}
\lim_{L\rightarrow 0}(\Delta
M)^{ren}_{\phi_{0}}=-m\left(\frac{L_{1}}{24L}\right)\ \ ;\ \
\lim_{L\rightarrow L_{1}}(\Delta M)^{ren}_{\phi_{0}}=-0.306m.
\end{equation}

For $L>L_{1}$, the vacuum solutions are $\phi_{\pm 1}$ with their
classical energies shown in Fig.~\ref{orbifoldfig1}. Since the
spectrum of perturbations in this case is not known analytically,
we cannot compute the quantum corrections to this energy
explicitly as we have done above for $\phi_{0}$. One way to
estimate the behavior of the quantum corrections to the energy of
$\phi_{\pm 1}$ is by calculating them as series in powers of $k$,
the modular parameter of the elliptic functions. As shown in
\cite{manton}, one can evaluate the eigenvalues of the
perturbations of $\phi_{\pm 1}$ in powers of $k$. To order
$k^{2}$, we have,
\begin{equation}
\omega_{p}^{2}=\left\{
\begin{array}{ll}
3k^{2}m^{2},&p=1\\
m^{2}(p^{2}-1)-\frac{3}{2}k^{2}m^{2}(p^{2}-2),&p=2,3,\dots
\end{array}
\right.
\end{equation}
Hence,
\begin{eqnarray}
(\Delta M)^{ren}_{\phi_{\pm 1}}&=&\left.(\Delta
M)^{ren}_{\phi_{0}}\right|_{L=L_{1}}+m\left(\frac{\sqrt{3}}{2}k\right)+O(k^{2})\nonumber\\
&=&m\left(-0.306+\frac{\sqrt{3}}{2}k+O(k^{2})\right)\label{corrections}
\end{eqnarray}
where only the term with $p=1$ contributes to $(\Delta
M)^{ren}_{\phi_{\pm 1}}$ to order $k$. From
Fig.~\ref{orbifoldfig3}, we can see that for $L<L_{1}$ the quantum
corrections to the vacuum energy decreases with $L$ as $L$
approaches $L_{1}$. On the other hand, from
Eq.~(\ref{corrections}), for $L>L_{1}$, the quantum corrections
increases with $k$ or $L$ as $L$ increases from $L_{1}$. Thus,
there is a discontinuity in the first derivative of the total
vacuum energies at $L=2\pi/m$ (Eq.~(\ref{critical})), indicating
again the presence of a phase transition that we have mentioned
before.

\section{Including fermions}

To include fermions in our model, we consider the
(1+1)-dimensional supersymmetric Lagrangian
\cite{divecchia,witten},
\begin{equation}
L=\frac{1}{2}(\partial_{\mu}\phi)^{2}-\frac{1}{2}W^{2}
+\frac{i}{2}\bar{\psi}\gamma^{\mu}\partial_{\mu}\psi-
\frac{1}{2}W'\bar{\psi}\psi,\label{susy}
\end{equation}
where the superpotential
\begin{equation}
W=\sqrt{\frac{\lambda}{2}}\left(\phi^{2}-\frac{m^{2}}{\lambda}\right)
\end{equation}
Here $\psi$ is a Majorana spinor. Note that the bosonic part is
the same as the model in the last section. This Lagrangian is
invariant under the supersymmetric transformation,
\begin{eqnarray}
\delta\phi&=&\bar{\epsilon}\psi\nonumber\\
\delta\psi&=&-(i\gamma^{\mu}\partial_{\mu}\phi+W)\epsilon,
\end{eqnarray}
with the corresponding supersymmetric current,
\begin{equation}
J^{\mu}=(\gamma^{\nu}\partial_{\nu}\phi+iW)\gamma^{\mu}\psi,
\end{equation}
and the supercharges,
\begin{eqnarray}
Q&=&\int dx J^{0}\nonumber\\ &=&\int dx
(\gamma^{\nu}\partial_{\nu}\phi+iW)\gamma^{0}\psi.
\end{eqnarray}
The supersymmetric algebra reads
\begin{equation}
Q_{1}^{2}=Q_{2}^{2}=2H\ \ ,\ \ \{Q_{1},Q_{2}\}=0
\end{equation}
where $H$ is the Hamiltonian, $Q=\left(\begin{array}{c} Q_{1}\\
Q_{2} \end{array} \right)$, and we have chosen the gamma matrices
$\gamma^{0}=\sigma_{2}$ and $\gamma^{1}=i\sigma_{3}$.

On the orbifold $S^{1}/Z_{2}$, we can see that under the
transformation,
\begin{equation}
\phi(-x)=-\phi(x)\ \ ,\ \ \psi(-x)=\pm\sigma_{3}\psi(x)
\end{equation}
the supersymmetric Lagrangian in Eq.~(\ref{susy}) is invariant.
Hence, we can choose the orbifold constraint for the fermionic
field as
\begin{equation}
\psi(-x)=\sigma_{3}\psi(x)\Rightarrow
\left(\begin{array}{c}\psi_{1}(-x)\\
\psi_{2}(-x)\end{array}\right)=\left(\begin{array}{c}\psi_{1}(x)\\
-\psi_{2}(x)\end{array}\right),\label{fermion}
\end{equation}
that is, $\psi_{1}$ is even and $\psi_{2}$ is odd. We could have
chosen a minus sign in Eq.~(\ref{fermion}) for the fermionic
field. This would only interchange the roles of $\psi_{1}$ and
$\psi_{2}$.

As in the bosonic case, the vacuum solution for $L<L_{1}$ is
$\phi_{0}$, while for $L>L_{1}$, they are $\phi_{\pm 1}$. The
energies of these solutions are all nonzero so supersymmetry is
broken by these vacua \cite{witten}.

To calculate the quantum corrections, we consider first $\phi_{0}$
for $0\geq L\geq L_{1}$. The bosonic spectrum is again given by
Eq.~(\ref{phi0spect}). For the fermionic perturbations
$u(x,t)=u(x)e^{-i\omega_{F}t}$, we have the equation of a massless
fermion,
\begin{eqnarray}
&&\frac{du_{1}}{dx}=-i\omega_{F}u_{2}\ \ ,\ \
-\frac{du_{2}}{dx}=i\omega_{F}u_{1}\nonumber\\
&\Rightarrow&-\frac{d^{2}u_{1}}{dx^{2}}=\omega_{F}^{2}u_{1}\ \ ,\
\ -\frac{d^{2}u_{2}}{dx^{2}}=\omega_{F}^{2}u_{2}.
\end{eqnarray}
Due to the orbifold constraint, $u_{1}$ must be even and $u_{2}$
must be odd. Therefore, only one component of the fermionic
perturbation, $u_{1}$, can develop a zero mode,
\begin{equation}
u_{1}\sim {\rm const}\label{phi0zeromode}
\end{equation}
while $u_{2}$ cannot. This is the same mechanism to obtain chiral
fermion on the fixed points of the orbifold in the
(1+4)-dimensional setting \cite{georgi}. Here in (1+1)-dimensions,
we have
\begin{equation}
\gamma^{1}u_{1}=u_{1}\ \ ,\ \ \gamma^{1}u_{2}=-u_{2}.
\end{equation}

For the positive modes, we have
\begin{equation}
u_{1}\sim {\rm cos}\frac{2\pi p}{L}x\ \ ;\ \ u_{2}\sim {\rm
sin}\frac{2\pi p}{L}x
\end{equation}
and eigenvalues
\begin{equation}
\omega_{F}^{2}=\frac{4\pi^{2}p^{2}}{L^{2}},\ \ p=1,2,\dots
\end{equation}
The quantum correction is thus
\begin{eqnarray}
(\Delta
M)^{SUSY,ren}_{\phi_{0}}&=&\frac{1}{2}\sum\omega_{B}-\frac{1}{2}\sum\omega_{F}\nonumber\\
&=&\frac{mL_{1}}{2L}\sum_{p=1}^{\infty}
\left\{\left[p^{2}-\left(\frac{L}{L_{1}}\right)^{2}\right]^{1/2}-p\right\}
\end{eqnarray}
Using the same mass renormalization procedure as in the bosonic
case in the last section, we see that the fermionic contribution
just cancels the Casimir energy term. Hence, we have
\begin{equation}
(\Delta M)^{SUSY,ren}_{\phi_{0}}=mf(L/L_{1}),
\end{equation}
where $f(L/L_{1})$ is the function in Fig.~\ref{orbifoldfig2}. The
divergence at $L=0$ is thus cured by the inclusion of fermions in
the supersymmetric Lagrangian.

Next we consider the quantum corrections to the vacuum solutions
$\phi_{\pm 1}$ for $L>L_{1}$. As discussed in the last section,
the equation of the bosonic perturbations is the Lam\'e equation.
For the fermionic perturbations, we have
\begin{equation}
\left(\frac{d}{dx}+\sqrt{2\lambda}\phi_{1}\right)u_{1}=-i\omega_{F}u_{2}\
\
,\ \
\left(-\frac{d}{dx}+\sqrt{2\lambda}\phi_{1}\right)u_{2}=i\omega_{F}u_{1}
\label{susyqm}
\end{equation}
Although these equations are not exactly solvable, the zero modes
can nevertheless be obtained simply as \cite{cooper}
\begin{equation}
u_{10}\sim e^{-\int^{x} \sqrt{2\lambda}\phi_{1}}\ \ ,\ \
u_{20}\sim e^{\int^{x} \sqrt{2\lambda}\phi_{1}}\label{zeromode}
\end{equation}
Both of these zero modes are even. If the orbifold constraint is
imposed, only $u_{10}$ survives. The situation is thus the same as
that for $\phi_{0}$ with $L<L_{1}$.

From Eq.~(\ref{susyqm}), we see that the eigenstates of $u_{1}$
and $u_{2}$ are related by
\begin{equation}
u_{2n}\sim \left(\frac{d}{dx}+\sqrt{2\lambda}\phi_{1}\right)u_{1n}
\end{equation}
Hence, $u_{2n}$ will be automatically parity odd if $u_{1n}$ is
parity even.

As in the bosonic case, we can consider the quantum corrections of
the fermions to the vacuum energy perturbatively when $k$ is
small. Here we see that the zero modes remain to have zero
energies for all values of $k$ (Eq.~(\ref{zeromode})). For the
higher modes, the corrections are at least of order $k^2$ by
direct perturbative calculations similar to the bosonic case.
Hence, the quantum corrections to $\omega_{F}$ are at least of the
order $k^{2}$, while the bosonic ones are of the order $k$ as
shown in Eq.~(\ref{corrections}). To the lowest order of $k$, we
finally have
\begin{eqnarray}
(\Delta M)^{SUSY,ren}_{\phi_{\pm 1
}}&=&m\left(f(1)+\frac{\sqrt{3}}{2}k+O(k^{2})\right)\nonumber\\
&=&m\left(-0.264+\frac{\sqrt{3}}{2}k+O(k^{2})\right)
\end{eqnarray}

\section{Conclusions and Discussions}

We have considered the vacuum structure of the (1+1)-dimensional
$\phi^4$ theory on the orbifold $S^{1}/Z_{2}$. When the size of
the orbifold is varied, we have found that a phase transition
occurs at $L=L_{c}=2\pi/m$. For $L<L_{c}$, there is a unique
classical vacuum solution $\phi_{0}=0$, while for $L>L_{c}$, there
are two degenerate vacua, $\phi_{1}$ and $\phi_{-1}$. It is worth
to note that this phase transition occurs only after imposing the
orbifold constraint together with the periodic boundary condition.
This phenomenon has been overlooked before because the
$L\rightarrow\infty$ limit is usually taken in constructing
orbifold GUT models. We think that this phase transition will be
important when one consider the case of dynamical compact
dimensions, especially in the cosmological setting in the early
universe.

We have also calculated the quantum corrections to the vacuum
solutions from the bosonic contributions using zeta-function
regularization to deal with the divergent quantities. As shown in
Fig.~\ref{orbifoldfig3}, the correction is dominated by the
Casimir energy for small $L$ and it goes to negative infinity as
$L\rightarrow 0$. On the other hand, as $L\rightarrow L_{c}$, the
correction decreases to a finite value. For $L>L_{c}$, we use
perturbative method to estimate the quantum correction. For $L$
close to $L_{c}$, the correction increases with $L$. Hence, the
quantum correction has a dip at $L=L_{c}$ with a discontinuity in
the slope, which is another indication of the presence of the
phase transition.

Fermions are included in the model by using a supersymmetric
Lagrangian. Since the vacuum solutions all have non-zero energies,
the supersymmetries, with supercharges $Q_{1}$ and $Q_{2}$, are
broken. With the fermions the main difference is that the Casimir
energy in the quantum corrections is cancelled by the fermionic
contributions. Then the correction goes to zero, instead of
negative infinity, as $L\rightarrow 0$.

The results obtained here should also be relevant to cases in
higher dimensions. For example, in the case of five dimensions
with one space compactified to an orbifold, the fields there can
be expressed as products of a four-dimensional part and another
part which depends only on the orbifold dimension \cite{georgi}.
Then one needs to consider the two different vacuum structures for
the size of the internal dimension smaller or larger than $L_{c}$.
Moreover, the fermionic zero modes (Eqs.~(\ref{phi0zeromode}) and
(\ref{zeromode})) are different in these two cases. In fact, they
have different forms around the fixed points at $x=0$ and $x=L/2$
which means that they could have different phenomenology on the
physical dimensions.

Other than going to higher dimensions, it is interesting to see
what the vacuum structures as well as the soliton solutions of the
theory when gauge fields are included on the orbifold. In
\cite{dermisek}, the monopole string solution, which is
independent of the compact dimension, is generalized to the
orbifold case. In that respect, one can also look at the
instanton, or caloron, solutions with or without non-trivial
holonomy, that is, whether there is symmetry breaking or not
\cite{gross,kraan}. We hope to look at these cases in the future
publications.

\begin{acknowledgments}
This work is supported by the National Science Council of the
Republic of China under contract number NSC 93-2112-M-032-008.
\end{acknowledgments}

\end{document}